\documentclass[twocolumn,amsmath,amssymb,showpacs,prl,aps,longbibliography,superscriptaddress]{revtex4-1}
\pdfoutput=1
\usepackage[bookmarks=false,linkcolor=blue,urlcolor=blue,colorlinks,citecolor=blue]{hyperref}
\usepackage{graphicx}
\usepackage{bm}
\usepackage{amssymb}
\usepackage{amsmath}
\usepackage{amsfonts}
\usepackage{nicefrac}
\usepackage{epsfig}
\usepackage{stackrel}
\usepackage{paralist}
\usepackage[rgb]{xcolor}
\usepackage{todonotes}
\usepackage{mathrsfs}
\usepackage{tabularx}
\newcommand{\eg}{\emph{e.g.}}
\newcommand{\ie}{\emph{i.e.}}

\newcommand{\wrt}{w.r.t.}

\def\multiset#1#2{\ensuremath{\left(\kern-.3em\left(\genfrac{}{}{0pt}{}{#1}{#2}\right)\kern-.3em\right)}}

\global\long\def\ket#1{\left|#1\right\rangle }

\newtheorem{corollary}{Corollary}
\newtheorem{theorem}{Theorem}

\newtheorem{claim}{Claim}

\def\be{\begin{equation}}
\def\ee{\end{equation}}
\def\beas{\begin{align*}}
\def\eeas{\end{align*}}
\def\bea{\begin{align}}
\def\eea{\end{align}}

\newcommand{\h}{{\mathbf h}}
\newcommand{\x}{{\mathbf x}}
\newcommand{\y}{{\mathbf y}}

\newcommand{\s}{{\mathbf s}}
\newcommand{\e}{{\mathbf e}}
\newcommand{\g}{{\mathbf g}}
\newcommand{\aaa}{{\mathbf a}}

\newcommand{\ff}{{\mathbf f}}

\newcommand{\A}{{\mathcal A}}

\newcommand{\R}{{\mathbb R}}
\newcommand{\N}{{\mathbb N}}

\newcommand{\abs}[1]{\left\lvert#1 \right\rvert}

\def\be{\begin{equation}}
\def\ee{\end{equation}}
\def\ba{\begin{align}}
\def\ea{\end{align}}
\newcommand{\bmat}{\begin{bmatrix}}
	\newcommand{\emat}{\end{bmatrix}}

\begin{document}
	\title{Quantum Entanglement in Deep Learning Architectures}
	\author{Yoav Levine}
	\email{yoavlevine@cs.huji.ac.il}
	\affiliation{The Hebrew University of Jerusalem, Israel}
	\author{Or Sharir}
	\email{or.sharir@cs.huji.ac.il}
	\affiliation{The Hebrew University 	of Jerusalem, Israel}
	\author{Nadav Cohen}
	\email{cohennadav@ias.edu}
	\affiliation{School of Mathematics, Institute for Advanced Study, Princeton, NJ, USA}
	\author{Amnon Shashua}
	\email{shashua@cs.huji.ac.il}
	\affiliation{The Hebrew University of Jerusalem, Israel}

	\begin{abstract}
		Modern deep learning has enabled unprecedented achievements in various domains. Nonetheless, employment of machine learning for wave function representations is focused on more traditional architectures such as restricted Boltzmann machines (RBMs) and fully-connected neural networks.
		In this letter, we establish that contemporary deep learning architectures, in the form of deep convolutional and recurrent networks, can efficiently represent highly entangled quantum systems. 
		By constructing Tensor Network equivalents of these architectures, we identify an inherent re-use of information in the network operation as a key trait which distinguishes them from standard Tensor Network based representations, and which enhances their entanglement capacity.
		Our results show that such architectures can support volume-law entanglement scaling, polynomially more efficiently than presently employed RBMs. 
		Thus, beyond a quantification of the entanglement capacity of leading deep learning architectures, our analysis formally motivates a shift of trending neural-network based wave function representations closer to the state-of-the-art in machine learning. 
	\end{abstract}
	\maketitle
	
	\emph{Introduction.--} 	
	Many-body physics and machine learning are distinct scientific disciplines, however they share a common need for efficient representations of highly expressive multivariate function classes. 
	In the former, the function class of interest captures the entanglement properties of examined many-body quantum systems, and in the latter, it describes the dependencies required for performing modern machine learning tasks.
	
	A prominent approach for classically simulating many-body wave functions makes use of their entanglement properties in order to construct Tensor Network (TN) architectures that aptly model them in the thermodynamic limit~\cite{fannes1992finitely,perez2007matrix,verstraete2004renormalization,vidal2008class,verstraete2008matrix,gu2009tensor,evenbly2011tensor,evenbly2014scaling}. 
	Though this method is successful in modeling one-dimensional (1D) systems that obey area-law entanglement scaling with sub-system size~\cite{eisert2010colloquium} through the Matrix Product State (MPS) TN~\cite{fannes1992finitely,perez2007matrix}, it still faces difficulties in modeling two-dimensional (2D) systems due to intractability~\cite{verstraete2004renormalization,orus2014practical}.
	
	In the seemingly unrelated field of machine learning, deep network architectures have exhibited an unprecedented ability to tractably encompass the convoluted dependencies that characterize difficult learning tasks such as image classification or speech recognition~\cite{NIPS2012_4824,simonyan2014very,Szegedy:2014tb,he2016deep,sutskever2011generating,graves2013speech,bahdanau2014neural,amodei2016deep}. 
	A consequent machine learning inspired approach for modeling wave functions makes use of fully-connected neural-networks and restricted Boltzmann machines (RBMs)~\cite{carleo2017solving,saito2017solving,PhysRevB.96.195145,gao2017efficient,PhysRevX.7.021021,carleo2018constructing,cai2018approximating}, which represent relatively veteran machine learning constructs. 
	
	\begin{figure}
		\centering
		\includegraphics[width=1\linewidth]{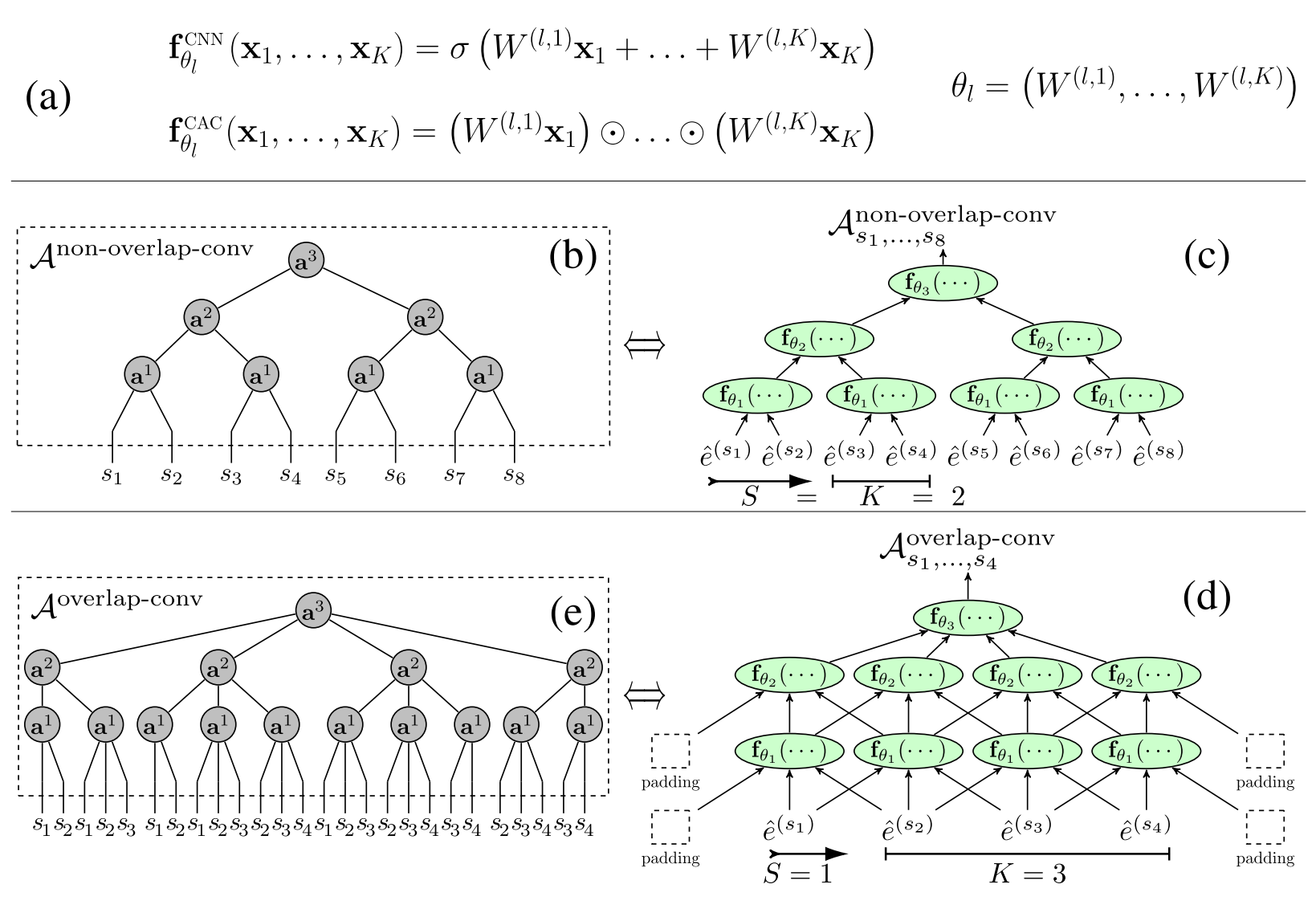}
		\caption{\label{fig:convacs} 
			Convolutional networks: In each layer $l\in\{1,...,L\}$ of a depth-$L$ convolutional network, convolution kernels of size $K$ are slid across the input maps, computing the layer outputs after every stride of $S$ steps.
			\textbf{(a)}
			CACs	and common CNNs share the above architectural description, and the type of convolutional network is determined by the function which computes the layers' outputs. The $\sigma(\cdot)$ function
			defining $\ff^{\textrm{CNN}}_{\theta_l}$ is some element-wise non-linearity in the form of a sigmoid or ReLU~\cite{glorot2011deep}, and the operation denoted $\odot$
			defining  $\ff^{\textrm{CAC}}_{\theta_l}$ stands for element-wise multiplication between
			vectors. 
			The function parameters are the convolution weights matrices $\theta_l=(W^{\textrm{(1,l)}},...,W^{\textrm{(K,l)}})$. 
			\textbf{(b,c)} The calculation of a non-overlapping convolutional network (for which $K=S$) assumes a tree structure. 
			A deep CAC with a $K=S$ restriction is equivalent to a Tree TN with internal order-($K+1$) tensors, obeying index-wise: $(\aaa^l)_{i,j_1,...,j_K}=W^{\textrm{(1,l)}}_{i,j_1},...,W^{\textrm{(K,l)}}_{i,j_K}$.
			\textbf{(d,e)} In overlapping-convolutional networks, which achieve state-of-the-art performance, the convolution kernel stride is $S=1$ (the output is calculated for every step of the kernel) and kernels are of general size, with typical values of $K=3$ or $K=5$. These values result in overlap of convolution kernels along the computation. The $0$-padding at the edges ensures that a layer's output is produced also for the edge activations.  The TN corresponding to the calculation of the		overlapping CAC must account for the inherent reuse of information due to the overlaps. Therefore, it involves duplication of external TN indices, such that re-used data is generated again and again by the
			TN, which cannot simply ‘copy-paste’ information. Thus, the
			portrayed recursive TN structure is received. 
			The presentation is in 1D form for clarity, and extensions to 2D are straightforward.
		}
	\end{figure}

	In this letter, we formally establish that highly entangled many-body wave functions can be efficiently represented by deep learning architectures that are at the forefront of recent empirical successes. 
	Specifically, we address two prominent architectures in the form of convolutional neural networks (CNNs), commonly used over spatial inputs (\eg~image pixels~\cite{NIPS2012_4824}), and recurrent neural networks (RNNs), commonly used over temporal inputs (\eg~phonemes of speech~\cite{amodei2016deep}).
	
\begin{figure}
		\centering
		\includegraphics[width=1\linewidth]{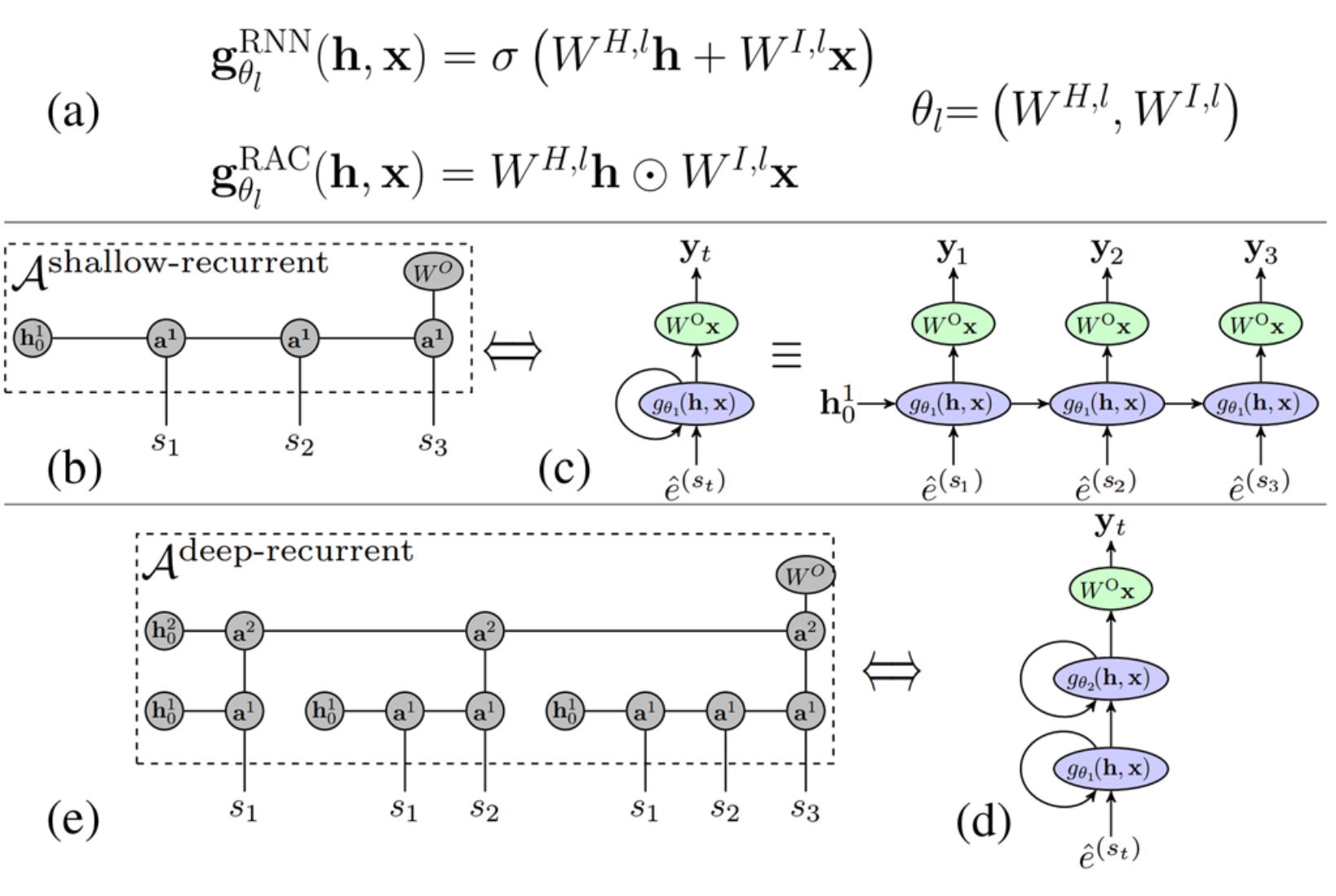}
		\caption{\label{fig:racs}
			Recurrent networks: In a recurrent network, an incoming input is integrated with an existing hidden state that the network computes from previous inputs. \textbf{(a)} RACs and common RNNs share the above architectural description, and the type of recurrent network is determined by the integration function.
			The function parameters are the hidden and input weights matrices $\theta_l=(W^{\textrm{(H,l)}},W^{\textrm{(I,l)}})$, and an output weights matrix $W^{{O}}$. 
			\textbf{(b,c)} The shallow RAC is equivalent to an MPS TN with internal order-$3$ tensors obeying index-wise: $\aaa^1_{ijk}=W^{\textrm{(H,1)}}_{ij}W^{\textrm{(I,1)}}_{ik}$. $\h_0^1$ is some initial hidden state.
			\textbf{(d,e)} A depth $2$ RAC is represented by a `recursive MPS' TN structure, which makes use of input duplication to circumvent an inherent inability of TNs to model information re-use. The TN in (e) is portrayed for the example of $N=3$.} 	
	\end{figure}

\begin{figure}
		\includegraphics[width=0.7\linewidth]{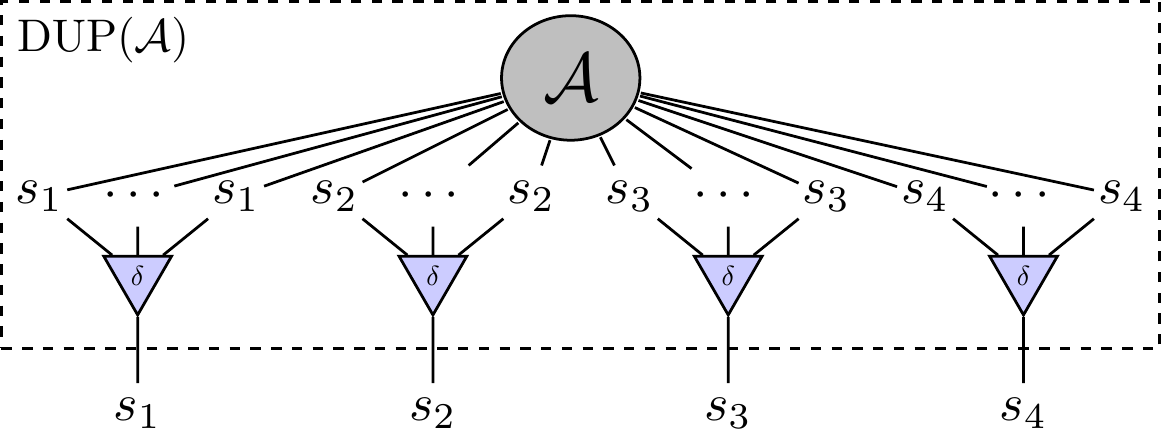}\vspace{-0.3cm}
		\caption{\label{fig:dup} Given a high-order tensor $\A$ with duplicated external indices [such as in Figs.\hyperref[fig:convacs]{~\ref{fig:convacs}(e)} and\hyperref[fig:racs]{~\ref{fig:racs}(e)}], the process of obtaining a dup-tensor $DUP(\A)$ involves a single $\delta$-tensor (hosting $1$ on the super-diagonal and $0$ otherwise) per unique external index.}
	\end{figure}

	The starting point of our analysis is a one-to-one equivalence established between a specialized version of these architectures 
	\footnote{Specialized in the sense that the architecture is the same and all non-linearities boil down to polynomials~\cite{cohen2016expressive,levine2017benefits}} and Tree/MPS TNs~\cite{levine2018deep,levine2017benefits}.
	We examine architectural extensions of the networks shown to be equivalent to common TNs, in the form of overlap of convolution kernels in CNNs and stacked layers in RNNs [see Figs.\hyperref[fig:convacs]{~\ref{fig:convacs}(d)} and\hyperref[fig:racs]{~\ref{fig:racs}(d)}]. These extensions are frequently used by machine learning practitioners in order to achieve state-of-the-art performance~\cite{NIPS2012_4824,graves2013speech}.	
	We identify in their structure an inherent re-use of information that cannot be naively represented in TN language, and is in fact a key factor that boosts the power of deep learning representations relative to common TNs. 
	Accordingly, by employing a TN manifestation of information re-use (see Fig.~\ref{fig:dup}), we construct TNs that incorporate state-of-the-art deep learning principles.
	
	We prove that convolutional networks can support volume-law entanglement scaling (Theorem~\ref{theorem:overlaps}), polynomially more efficiently in resources compared with RBM based representations in 2D~\cite{carleo2017solving,PhysRevX.7.021021}. Furthermore, we prove that recurrent networks can support logarithmic corrections to the area-law entanglement scaling in 1D (Theorem~\ref{theorem:deep_racs}).
	We thus assert formal benefits of using state-of-the-art deep learning principles for the investigation of highly entangled many-body systems. 	 
	
	\emph{Entanglement in convolutional networks.--}
	Representing an $N$-particle wave function through any computational scheme amounts to representing a corresponding order-$N$ tensor, $\A_{s_1\cdots s_N}$. Here, the indices $s_j$ for $j\in\{1,\ldots,N\}$ run in $\{1,\ldots,M\}$, where $M$ is the dimension of the local Hilbert space of each particle. 
	The algebraic approach manifested by TNs is a compact representation of the higher-order tensor through contractions among lower-order tensors. 
	For the example of the Tree TN depicted in Fig.\hyperref[fig:convacs]{~\ref{fig:convacs}(b)}, a high order tensor is attained by hierarchically arranging order-$3$ tensors. 
	
	Below, we investigate the entanglement scaling that can be supported by deep convolutional networks, a successful class of deep learning architectures leading empirical breakthroughs in image processing and more~\cite{NIPS2012_4824,simonyan2014very,szegedy2016rethinking}.
	Ref.~\cite{levine2018deep} has identified a Tree TN with the operation of a deep convolutional network referred to as a Convolutional Arithmetic Circuit (CAC)~\cite{cohen2016expressive,cohen2016convolutional,cohen2017inductive}, see Fig.\hyperref[fig:convacs]{~\ref{fig:convacs}(b,c)}.
	In the established mapping the order-$3$ tensors $\aaa^{l}\in\R^{r_{l-1}\times r_{l-1}\times r_{l}}$, comprising a tree node of level $l$ in Fig.\hyperref[fig:convacs]{~\ref{fig:convacs}(b)}, are described in machine learning language in terms of functions $\ff_{\theta_l}:\R^{r_{l-1}\times r_{l-1}}\to\R^{r_{l}}$, as depicted in Fig.\hyperref[fig:convacs]{~\ref{fig:convacs}(c)}. 
	Here, $\theta_l$ stands for the learned parameters defining the convolutional network computation in layer $l$ [see Fig.\hyperref[fig:convacs]{~\ref{fig:convacs}(a)}], and $r_l$ are the number of feature maps in layer $l$.
	CACs share the architectural features of standard CNNs, and they work well in many practical settings~\cite{sharir2016tractable,cohen2016deep,khrulkov2018expressive}. They differ from common networks by possessing a theoretically appealing trait: all non-linearities in $\ff_{\theta_l}$ boil down to polynomials~\cite{cohen2016expressive}. We will harness this trait for analyzing the achievable entanglement scaling of the wave functions represented by such networks.
	
	Two basic architectural traits defining the operation of a deep convolutional network are the kernel size $K$, denoting the spatial extent of the convolution window, and the stride size $S$, denoting the spatial distance between windows. For example, for $S=2$, the convolution kernel is slid in increments of $2$. 
	The above equivalence to TNs was not obtained for general CAC networks, but rather for a simplified version for which the convolution stride size $S$ is equal to the convolution kernel size $K$. In such $S=K$ networks, kernels of adjacent calculations do not overlap [see Fig.\hyperref[fig:convacs]{~\ref{fig:convacs}(c)}], and therefore they are referred to as non-overlapping convolutional networks. 
	This restriction yields a tree structured calculation that has enabled the established equivalence to the Tree TN.
	
	In order to represent an order-$N$ tensor with a deep learning architecture, we consider network inputs comprised of $N$ standard-basis vectors of $\R^M$, $\{\hat{e}^{(s_j)}\}_{j=1}^N$ ($\hat{e}^{(s_j)}$ hosts $1$ in the $s_j$'th position and $0$ otherwise, where $s_j\in\{1,\ldots,M\}$). 
	For such inputs, the output $y$ of a non-overlapping ($K=S$) CAC is the corresponding tensor entry:
	\begin{align}\label{eq:non_over_convac}
	y\left(\hat{e}^{(s_1)},...,\hat{e}^{(s_N)}\right)=\A^{\textrm{non-overlap-conv}}_{s_1...s_N},
	\end{align}
	where $\A^{\textrm{non-overlap-conv}}$ can be efficiently written as a Tree TN~\cite{levine2018deep}. 
	
	Per Eq.~\eqref{eq:non_over_convac}, the above network effectively  represents an $N$-particle quantum state denoted: {\footnotesize$\ket{\psi^{~\textrm{non-overlap-conv}}}:=\sum_{s_1,...,s_N=1}^M \A^{\textrm{non-overlap-conv}}_{s_1...s_N}\left|\hat{\psi}_{s_1...s_N}\right\rangle$}, where {\footnotesize$\left\{\left|\hat{\psi}_{s_1...s_N}\right\rangle\right\}_{s_1,...,s_N=1}^M$} is some orthonormal basis of the many-body Hilbert space. 
	The description of {\footnotesize$\ket{\psi^{~\textrm{non-overlap-conv}}}$} in terms of a Tree TN reveals an upper bound on the maximum entanglement capacity of the represented wave function: non-overlapping convolutional networks can support up to logarithmic corrections to area-law entanglement scaling in 1D and up to area-law entanglement in 2D~\footnote{These upper bounds on entanglement can be attained by employing minimal cut considerations in TNs~\cite{cui2016quantum}.}.
	
	A popular enhancement to Tree TNs is the Multiscale Entanglement Renormalization Ansatz (MERA) TN~\cite{vidal2008class}, which is known to be superior in modeling critical systems~\cite{vidal2007entanglement}. 
	The MERA TN introduces loops via the disentangling operations, which entail intractability of straightforward wave function amplitude computation~\footnote{Direct sampling of spin configuration from MERA was shown to be possible~\cite{ferris2012perfect,ferris2012variational}.}.
	Non-overlapping convolutional networks, which employ the same naive coarse-graining scheme of a tree, have prevalent machine learning enhancements that allow for an efficient computation of wave function amplitudes, enabling their optimization via stochastic sampling techniques such as in Ref.~\cite{carleo2017solving}.  
	Below we examine this enhancement and construct its corresponding TN, which differs from MERA, illuminating the competing mechanisms effectively elected by the many-body physics and deep learning communities for enhancing tree-like decimation schemes. 
	
	Specifically, state-of-the-art deep convolutional networks do not uphold the above $S=K$ restriction, but rather make use of size $K>1$ convolution kernels with stride $S=1$~\cite{NIPS2012_4824,simonyan2014very}, \ie~sliding the convolution kernels continuously (increments of $1$) during a layer's computation. 
	This architectural trait, which implies that the kernels overlap during computation [see Fig.\hyperref[fig:convacs]{~\ref{fig:convacs}(d)}], was shown to yield an exponential enhancement in network expressivity~\cite{sharir2018expressive} despite admitting a mere linear growth in the amount of parameters and in computational cost.
	
	The construction of a TN that matches the calculation of an overlapping CAC is less trivial than that of the non-overlapping case, since due to the overlaps the output vector of each layer is duplicated and used for computing several adjacent inputs of the subsequent layer.
	This inherent re-use of data, which is
	graphically represented by the multiple edges emanating out of the computational nodes in Fig.\hyperref[fig:convacs]{~\ref{fig:convacs}(d)}, and is 
	simply achieved in practice, is impossible to represent in the framework of TNs (see Supplemental Material for the formalization of this argument). 
	However, we obtain the form of a TN representing the overlapping-convolutional network with a simple `trick'~--~
	duplication of the input data itself, such that each instance of a duplicated intermediate vector is generated by a separate TN branch.
	This technique yields the `recursive-Tree' TN construction in Fig.\hyperref[fig:convacs]{~\ref{fig:convacs}(e)}, where for standard-basis input vectors the input duplication amounts to external index duplication.
	
	Due to these external duplications, the tensor represented by the TN in Fig.\hyperref[fig:convacs]{~\ref{fig:convacs}(e)}, denoted $\A^{\textrm{overlap-conv}}$, does not correspond to an $N$-particle wave function, since it has more than $N$ external edges ($N=4$ in the depicted example). 
	However, when considering the operation of the overlapping-convolutional network over inputs comprised of standard-basis vectors, $\{\hat{e}^{(s_j)}\}_{j=1}^N$, we may write the function realized by the network as: 
	\begin{align}\label{eq:overlaps}
	y\left(\hat{e}^{(s_1)},...,\hat{e}^{(s_N)}\right)=\left[ DUP(\A^{\textrm{overlap-conv}})\right]_{s_1...s_N},
	\end{align}
	where \emph{DUP}$(\A^{\textrm{overlap-conv}})$ is the order-$N$ sub-tensor of $\A^{\textrm{overlap-conv}}$ holding its values when duplicated external indices are equal. 
	Fig.~\ref{fig:dup} shows the TN calculation of \emph{DUP}$(\A)$, for a general tensor $\A$ with duplicated external indices, which employs $\delta$-tensors that operate similarly to the copy-tensors in Ref.~\cite{biamonte2011categorical} (a $\delta$-tensor has $1$ on the super-diagonal and $0$ otherwise). 
	
	Eq.~\eqref{eq:overlaps} implies that the overlapping CAC efficiently computes amplitudes of the state {\footnotesize$\ket{\psi^{\textrm{overlap-conv}}}:=\sum_{s_1,..,s_N=1}^M \left[ DUP(\A^{\textrm{overlap-conv}})\right]_{s_1...s_N}\left|\hat{\psi}_{s_1...s_N}\right\rangle$}.
	In the following, we rely on results regarding expressiveness of overlapping-convolutional architectures~\cite{sharir2018expressive} and bound from below the maximum entanglement of a state modeled by such networks:
	\begin{theorem}\label{theorem:overlaps} (Proof in Supplemental Material)\\
		Let {\footnotesize$\ket{\psi^{~\textnormal{\textrm{overlap-conv}}}}$} be the state represented by an overlapping CAC (stride $S=1$) of depth $L$ with convolution kernel size $K^d$ in $d$ spatial dimensions for $d=1,2$ .
		Let $(A,B)$ be a partition of $\{1,\ldots,N\}$ such that $A$ is  of size $\alpha$ in 1D and $\alpha \times \alpha$ in 2D ($\alpha$ is the linear dimension of the sub-system $A$), with $\abs{A}\leq\abs{B}$.
		Then, the maximal entanglement entropy \wrt~$(A,B)$ of {\footnotesize$\ket{\psi^{~\textnormal{\textrm{overlap-conv}}}}$} upholds:
		$$maxEE\left(\ket{\psi^{~\textnormal{\textrm{overlap-conv}}}}\right)=\Omega\left({\textrm{\small{min}}}\left\{\alpha^{d},LK\alpha^{d-1}\right\}\right).$$
	\end{theorem}
	Thus, an overlapping-convolutional network with $L$ layers
	supports volume-law entanglement scaling~--~$\alpha^d$, for systems of linear size $\alpha<LK$. 
	Practically, overlapping-convolutional networks with common characteristics of \eg~kernel size $K=5$ and depth $L=20$, can support the entanglement of any 2D system of interest up to sizes $100\times100$, which are unattainable by competing intractable approaches~\cite{gull2013superconductivity,chen2013evolution,lubasch2014algorithms,zheng2016ground,liu2017gradient,leblanc2015solutions}.
	
	Moreover, the result in Theorem~\ref{theorem:overlaps} indicates a significant advantage in modeling volume-law entanglement scaling of deep convolutional networks relative to competing veteran neural-network based approaches. 
	These approaches are promising to grant tractable access to 2D systems that cannot be modeled by 2D TNs, since they render computation of wave function amplitudes tractable and therefore stochastic-sampling based optimization techniques can be employed even for large 2D systems. 
	However, in order to represent volume-law entanglement in a 2D system of $N$ particles, fully-connected networks require an amount of network parameters that scales like $O(N^2)$~\cite{saito2017solving,cai2018approximating}, while RBMs require $O(N)$ parameters~\cite{carleo2017solving,PhysRevX.7.021021}. In contrast, since the amount of parameters in 2D overlapping-convolutional networks is proportional to $LK^2$, the above volume law condition $\alpha<LK$	implies the following corollary of Theorem~\ref{theorem:overlaps}:
	\begin{corollary}\label{corr:efficiency}
		The number of overlapping-convolutional network parameters required for modeling volume-law entanglement scaling in a 2D system of  $N$-particles, scales like $O(\sqrt{N})$.
	\end{corollary}
	Therefore, these networks have a clear polynomial advantage in resource efficiency over previously used fully-connected networks and RBMs.
	Thus, deep convolutional networks have the potential to provide access to highly-entangled 2D systems of sizes unattainable by such competing veteran neural-network based approaches. 
	
	\emph{Entanglement in recurrent networks.--}
	In the following, we investigate the entanglement scaling that can be supported by another successful class of deep learning architectures -- recurrent networks, responsible for empirical successes in machine translation~\citep{bahdanau2014neural}, speech recognition~\citep{graves2013speech,
		amodei2016deep}, and more.
	Ref.~\cite{levine2017benefits} has identified an MPS TN [Fig.\hyperref[fig:racs]{~\ref{fig:racs}(b)}] with the operation of a recurrent network referred to as a Recurrent Arithmetic Circuit (RAC) [Fig.\hyperref[fig:racs]{~\ref{fig:racs}(c)}]. 
	In the shown mapping, the order-$3$ tensor $\aaa^1\in\R^{R\times R\times M}$, which composes the translationally invariant MPS, is described in machine learning language in terms of a function $\g_{\theta_1}:\R^{R\times M}\to\R^{R}$. Here, $\theta_1$ stands for the learned parameters defining the recurrent network computation, $R$ is the number of hidden feature maps, and $M$ is the number of input feature maps.
	RACs work well in practice~\cite{khrulkov2018expressive} and share the architectural features of conventional RNNs, with polynomial non-linearities in $\g_{\theta_1}$.
	
	For standard-basis inputs, the output $y$ of a single-layered (shallow) RAC, depicted in Fig.\hyperref[fig:racs]{~\ref{fig:racs}(c)}, is given by:
	\begin{align}\label{eq:shallow_racs}
	y\left(\hat{e}^{(s_1)},...,\hat{e}^{(s_N)}\right)=\A^{\textrm{shallow-recurrent}}_{s_1...s_N},
	\end{align}
	where $\A^{\textrm{shallow-recurrent}}$ can be efficiently written as an MPS TN~\cite{levine2017benefits}. 
	Thus, the above network represents an $N$-particle quantum state in 1D, denoted: {\footnotesize$\ket{\psi^{~\textrm{shallow-recurrent}}}:=\sum_{s_1,...,s_N=1}^M \A^{\textrm{shallow-recurrent}}_{s_1...s_N}\left|\hat{\psi}_{s_1...s_N}\right\rangle$}, which can support no higher than area-law entanglement scaling.

	Similar to the convolutional network case, we extend the analysis to include state-of-the-art mechanisms in recurrent networks. An empirically successful architectural choice~\cite{graves2013speech,hermans2013training}, recently proven to bring forth a substantial advantage in a recurrent network long-term memory capacity~\cite{levine2017benefits}, involves adding more layers, \ie~deepening [see Fig.\hyperref[fig:racs]{~\ref{fig:racs}(d)}].
	Importantly, since the output vector of each layer of the deep RAC at every time-step is
	used twice 
	(as an input of the next layer up, but also as a hidden vector for the next time-step), 
	there is an inherent re-use of data during network computation.
	Therefore, we duplicate the inputs as in the overlapping-convolutional network case, and obtain the TN of the deep RAC, denoted $\A^{~\textrm{deep-recurrent}}$ [Fig.\hyperref[fig:racs]{~\ref{fig:racs}(e)}].
	
	The network operation over standard-basis inputs upholds: 
	\begin{align}\label{eq:deep_racs}
	y\left(\hat{e}^{(s_1)},...,\hat{e}^{(s_N)}\right)=\left[DUP(\A^{~\textrm{deep-recurrent}})\right]_{s_1...s_N},
	\end{align}
	where, as before, \emph{DUP}$(\A^{~\textrm{deep-recurrent}})$ is the order-$N$ sub-tensor of $\A^{~\textrm{deep-recurrent}}$ holding its values when duplicated external indices are equal (as illustrated in Fig.~\ref{fig:dup}).
	Relying on results regarding expressiveness of deep recurrent architectures~\cite{levine2017benefits}, we examine the entanglement scaling of a represented state {\footnotesize$\ket{\psi^{~\textrm{deep-recurrent}}}:=\sum_{s_1,..,s_N=1}^M \left[DUP(\A^{~\textrm{deep-recurrent}})\right]_{s_1...s_N}\left|\hat{\psi}_{s_1...s_N}\right\rangle$} [Eq.~\eqref{eq:deep_racs}]:
	\begin{theorem}\label{theorem:deep_racs}(Proof in Supplemental Material)\\
		Let $(A,B)$ be a partition of $\{1,\ldots,N\}$ such that $\abs{A}\leq\abs{B}$ and $B=\{1,\ldots,\abs{B}\}$~\footnote{We focus on the case where $A$ is located to the right $B$ for proof simplicity, simulations of the network in Fig.\hyperref[fig:racs]{~\ref{fig:racs}(d)} with randomized weights matrices indicate that the lower bound in Theorem~\ref{theorem:deep_racs} holds for all other locations of $A$.}. 
		Then, the maximal entanglement entropy \wrt~$(A,B)$ of {\footnotesize$\ket{\psi^{\textnormal{\textrm{deep-recurrent}}}}$} upholds: $$maxEE\left(\ket{\psi^{\textnormal{\textrm{deep-recurrent}}}}\right)=\Omega\left(log\left\{\abs{A}\right\}\right).$$
	\end{theorem} 
	Thus, the deep-recurrent network in Fig.\hyperref[fig:racs]{~\ref{fig:racs}(d)}
	supports logarithmic corrections to the area-law entanglement scaling with sub-system size in 1D, similarly to the MERA TN.
	As in the convolutional network case, we see that deep learning practitioners employ enhancements that admit only a linear growth in parameters and computation time (adding recurrent layers), but at the same time introduce a considerable boost to network expressivity.
	The above attained entanglement scaling in successful deep recurrent architectures hints at a critical nature of data sets on which they are employed, such as text, speech, etc.
	
	\emph{Discussion.--} 
	Previous works have employed TN equivalents of simplified convolutional networks (non-overlapping) and recurrent networks (single-layered), and lent well-established many-body physics concepts for principled  deep learning architecture design~\cite{levine2018deep,levine2017benefits}. 
	Above, we have shown how the introduction of  performance enhancing mechanisms that are commonly employed by deep learning practitioners, entails wave function representations which cannot be matched by common TNs.
	Our work quantifies the power of deep learning for highly-entangled wave function representations, theoretically motivating a shift towards employment of state-of-the-art deep learning architectures in many-body physics research.
	
	Beyond the shown polynomial efficiency of convolutional networks in resources required for representing 2D volume-law entanglement (Corollary~\ref{corr:efficiency}), they enjoy an advantage relative to previously employed RBMs in the optimization aspect as well. 
	The cost of computing a wave function amplitude per spin configuration [Eq.~\eqref{eq:overlaps}] is linear in system size for convolutional networks versus quadratic for used RBMs.
	Moreover, the resource efficiency in Corollary~\ref{corr:efficiency}, together with the locality of the convolution computation relative to long-ranged calculations used in RBMs to attain volume-law entanglement, can provide a further boost in optimization speed~--~there are significant non-linear runtime advantages to having low memory together with local operations when employing Graphical Processing Units, which have substantially sped up deep networks in recent years. 
	Overall, we believe that the results presented in this letter can help bring quantum many-body physics and state-of-the-art machine learning approaches one step closer together.
	
	\vspace{1mm}
	We thank Guifre Vidal, Thomas Spencer, John Imbrie, Bartlomiej Czech, Eyal Bairey, Eyal Leviatan, Miles Stoudenmire, Giuseppe Carleo, Ivan Glasser and Lei Wang for useful discussions. This work is supported by ISF Center grant 1790/12 and by the
	European Research Council (TheoryDL project). Nadav Cohen is a member of the Zuckerman Israeli Postdoctoral
	Scholars Program, and is supported by the Schmidt Foundation. Yoav Levine is supported by the Adams Fellowship Program of the
	Israel Academy of Sciences and Humanities.
	\vspace{1mm}
	\appendix
	\section{A.``No-Cloning" in Tensor Networks}
	The required
	operation of duplicating a vector and sending it to be part of two different
	calculations, which is simply achieved in any practical setting, is actually
	impossible to represent in the framework of TNs. We formulate this notion in the
	following claim:
	\begin{claim} \label{claim:no_clone}
		Let $v\in\R^P,P\in\N$ be a vector. $v$ is represented by a node with one leg
		in the TN notation. The operation of duplicating this node, \ie~ forming two
		separate nodes of degree $1$, each equal to $v$, cannot be achieved by any
		TN.
	\end{claim}
	\emph{Proof.}
	We assume by contradiction that there exists a TN $\phi$ which operates on any vector $v\in\R^P$ and clones it to two separate nodes of degree $1$, each equal to $v$, to form an overall TN representing $v\otimes v$.
	Component wise, this implies that $\phi$ upholds $\forall v\in\R^P:~\sum_{i=1}^P \phi_{ijk}v_i=v_jv_k$.
	By our assumption, $\phi$ duplicates the standard basis elements of $\R^P$, denoted $\{\hat{\e}^{(\alpha)}\}_{\alpha=1}^P$, meaning that $\forall \alpha\in[P]$:
	\begin{equation}\label{eq:no_clone}
	\sum_{i=1}^P \phi_{ijk}\hat{e}^{(\alpha)}_i=\hat{e}^{(\alpha)}_j\hat{e}^{(\alpha)}_k.
	\end{equation}
	By definition of the standard basis elements, the left hand side of Eq.~\eqref{eq:no_clone} takes the form $\phi_{\alpha jk}$ while the right hand side equals $1$ only if $j=k=\alpha$, and otherwise $0$. In other words, in order to successfully clone the standard basis elements, Eq.~\eqref{eq:no_clone} implies that $\phi$ must uphold $\phi_{\alpha jk}=\delta_{\alpha jk}$. However, for $v=\mathbf{1}$, \ie~$\forall j\in[P]:~v_j=1$, a cloning operation does not take place when using this value of $\phi$, since $\sum_{i=1}^P \phi_{ijk}v_i=\sum_{i=1}^P \delta_{ijk}=\delta_{jk}\neq 1= v_iv_j$, in contradiction to $\phi$ duplicating any vector in $\R^P$.
	
	\hfill $\square$ 
	
	\section{B. Entanglement Scaling in Overlapping-Convolutional Networks}
	
	In this section we provide a detailed description of overlapping convolutional networks that include spatial decimation, and then provide our proof of Theorem~1 of the main text. Additionally, we analyze the effect of the added decimating pooling layers on the entanglement entropy.
	
\begin{figure}
		\centering
		\includegraphics[width=\linewidth]{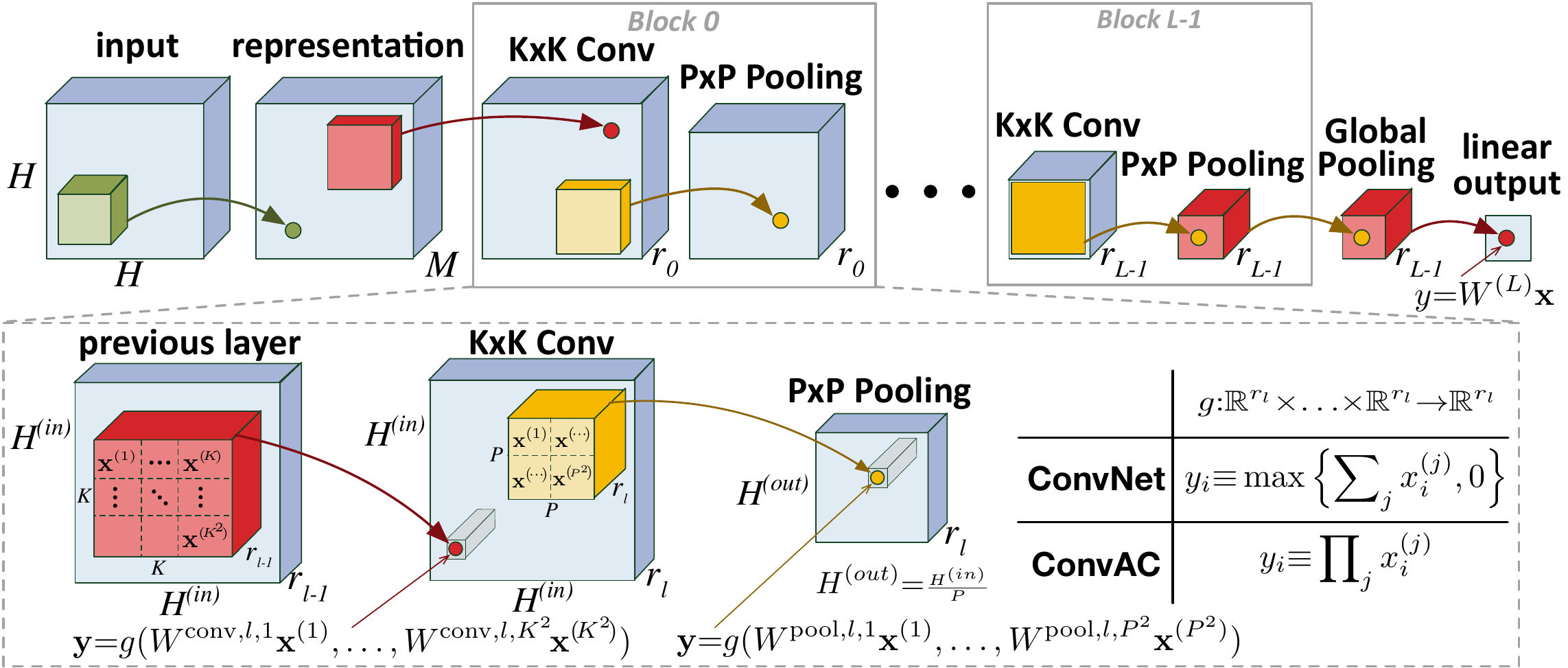}
		\caption{\label{fig:general_overlaps} A general overlapping convolutional network in 2D.}
	\end{figure}

	We begin by presenting a broad definition of a what is called a Generalized Convolutional~(GC)
	layer~\citep{sharir2018expressive} as a fusion of a $1{\times}1$ linear operation with a pooling (spatial decimation)
	function~--~this view of convolutional layers is motivated by the
	all-convolutional architecture~\citep{springenberg2015striving}, which replaces all
	pooling layers with convolutions with stride greater than 1. The input to a
	GC layer is an order 3 tensor, having width
	and height equal to $H^{(\text{in})} \in \N$ and depth $r^{(\text{in})} \in \N$,
	also referred to as channels, e.g. the input could be a 2D image with RGB color
	channels. Similarly, the output of the layer has width and height equal to
	$H^{(\text{out})} \in \N$ and $r^{(\text{out})} \in \N$ channels, where
	$H^{(\text{out})} = \frac{H^{(\text{in})}}{S}$ for $S \in \N$ that is referred
	to as the \emph{stride}, and has the role of a sub-sampling operation. Each
	spatial location $(i,j)$ at the output of the layer corresponds to a 2D window
	slice of the input tensor of size $R \times R \times r^{(\text{in})}$, extended
	through all the input channels, whose top-left corner is located exactly at
	$(i\cdot S, j\cdot S)$, where $R \in \N$ is referred to as its \emph{local
		receptive field}, or filter size. For simplicity, the parts of window slices
	extending beyond the boundaries have zero value. Let
	$\y \in \R^{r^{\text(out)}}$ be a vector representing the channels at some
	location of the output, and similarly, let
	$\x^{(1)},\ldots,\x^{(R^2)} \in \R^{r^{(\text{in})}}$ be the set of vectors
	representing the slice, where each vector represents the channels at its
	respective location inside the $R \times R$ window, then the operation of a
	GC layer is defined as follows:
	\begin{equation*}
	\y = g(W^{(1)}\x^{(1)}, \ldots, W^{(R^2)}\x^{(R^2)}),
	\end{equation*}
	where $W^{(1)},\ldots,W^{(R^2)} \in \R^{r^{(out)} \times r^{(in)}}$ are referred to as the
	weights of the layer, and
	$g:\R^{r^{(out)}} \times \cdots \times \R^{r^{(out)}} \to \R^{r^{(out)}}$ is
	some point-wise pooling function. Additionally, we call a GC layer that is limited to unit-stride and has
	$K{\times}K$ receptive field a \emph{$K{\times}K$ Conv layer}, and similarly, a
	\emph{$P{\times}P$ Pooling layer} is a GC layer with both stride and receptive
	fields equal to $P{\times}P$. With the above definitions, a convolutional network is simply a
	sequence of $L$ blocks of Conv and Pooling layers that follows the representation layer, 
	and ends with a global pooling layer, i.e. a pooling layer with $P$ equals the entire spatial
	extent of its input. The entire network is illustrated in Fig.~\ref{fig:general_overlaps}.
	
	Given a non-linear point-wise activation function $\sigma(\cdot)$ (e.g. ReLU), then setting all
	pooling functions to average pooling followed by the activation, i.e.
	$g(\x^{(1)},\ldots,\x^{(R^2)})_c=\sigma\left(\sum_{i=1}^{R^2} x^{(i)}_c\right)$
	for $c \in [D^{(\text{out})}]$,
	give rise to the common all-convolutional network with $\sigma(\cdot)$
	activations, which served as the initial motivation for this formulation.
	Alternatively, choosing instead a product pooling function, i.e.
	$g(\x^{(1)},\ldots,\x^{(R^2)})_c = \prod_{i=1}^{R^2} x^{(i)}_c$ for
	$c \in [D^{(\text{out})}]$, results in an Arithmetic Circuit, i.e.
	a circuit containing just product and sum operations, hence it is referred to
	as an Overlapping Convolutional Arithmetic Circuit, or Overlapping ConvAC in short,
	where `overlapping' refers to having receptive fields which overlap when $K > 1$. 
	The non-overlapping case, where $K = 1$, is equivalent to ConvACs as originally
	introduced by \citet{cohen2016expressive}.
	
	In the body of the paper we have discussed the entanglement entropy of overlapping convolutional
	networks with no spatial decimation, which essentially amount to having pooling layers with $P = 1$, which was summarized
	in Theorem~1 of the main text. The following theorem quantifies the effect of pooling layers with
	$P = 2$ in overlapping convolutional networks: 
	
	\begin{theorem}\label{theorem:overlaps_pool}
		For an overlapping ConvAC with $2^d$ pooling operations in between convolution layers (Fig.~\ref{fig:general_overlaps} with $P=2$), the maximal entanglement entropy \wrt~$(A,B)$ modeled by the network obeys:
		\begin{equation*}
		\Omega\left(\min\left\{\alpha^{d},K\alpha^{d-1}\right\}\right),
		\end{equation*}
		where $\alpha$ is the linear dimension of the d-dimensional system for $d=1,2$. 
	\end{theorem}
	
	Thus, the introduction of such pooling layers results in a diminished ability of the overlapping-convolutional network to represent volume-law entanglement scaling, since the $KL$ factor from Theorem~1 of the main text is diminished to a factor of $K$. 
	In the following, we prove the results in Theorem~1 of the main text and Theorem~\ref{theorem:overlaps_pool} in this appendix regarding entanglement scaling supported by overlapping ConvACs: 
	
	\emph{Proof} (of Theorem~1 of the main text and Theorem~\ref{theorem:overlaps_pool} above).
	We begin by providing a succinct summary of the theoretical analysis of overlapping ConvACs that was shown by \cite{sharir2018expressive},
	including the necessary technical background on ConvACs required to understand their results. \cite{sharir2018expressive} shows lower
	bounds on the rank of the dup-tensor for various architectures when $A$ is left half of the input and $B$ the right half, in
	$d=2$, when the convolutional kernel is anchored at the corner instead of at the center like presented in this letter.
	
	\begin{figure}[h]
		\centering
		\includegraphics[width=\linewidth]{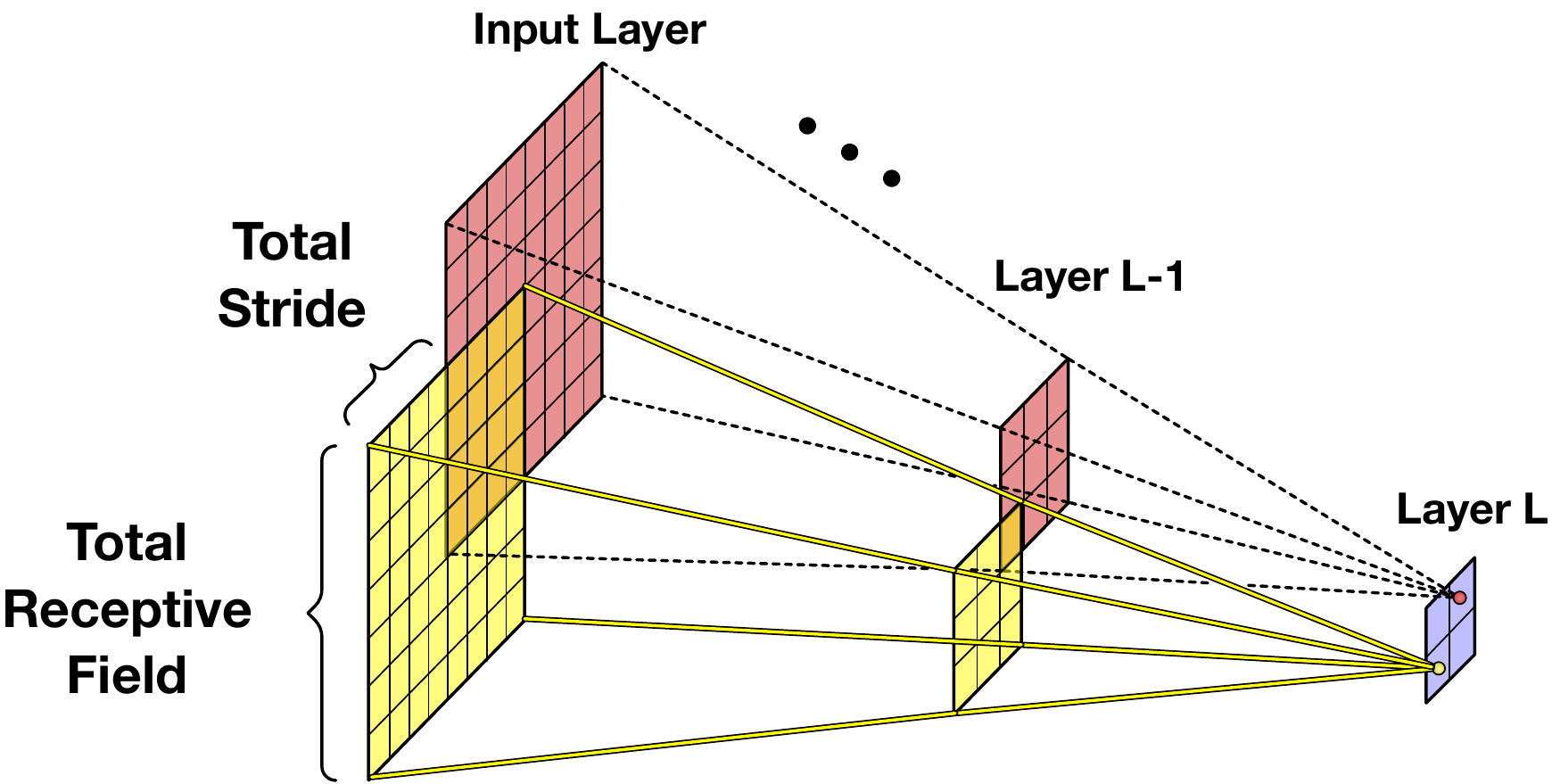}
		\caption{\label{fig:total_properties}Illustration of the total receptive field and the total stride.}
	\end{figure}
	
	For any layer $l\in[L]$ in a convolutional network, the local receptive field (or kernel size) $K^{(l)}$ is defined as the linear size of the window on
	which each convolutional kernel acts upon, and the stride $S^{(l)}$ is defined as the step size in each dimension between two neighboring windows (assumed to be $1$ in this letter). 
	The main result of \cite{sharir2018expressive} relies on two architecture dependent attributes that they referred to as the total receptive field and and
	the total stride of the $l$'th layer, defined as the projections on the input layer of the local receptive fields and strides from the perspective of the
	$l$'th layer, as illustrated in Fig.~\ref{fig:total_properties}.  In their main result they show that the first layer that has a total
	receptive field of at least half the linear dimension of the input size $N$, denoted $l_0$, gives rise to a lower bound on the rank of the matricized tensor that is
	proportional to $\min\{M,r_0,\ldots,r_{L-1}\}^{\mathcal{O}(\nicefrac{N^2}{4\cdot T^{(l_0)}_S})}$, where $T^{(l_0)}_S$ is the total stride of the $l_0$'th
	layer. To prove this result the authors rely on the ability of a sufficiently large total receptive field to represent identity matrices
	between pairs of input indices, each pair comprising one index from $A$ and one from $B$, where the total stride limits the maximal number of pairs as it denotes the minimal distance between any two given pairs of indices.
	
	To prove the lower bounds on the architectures described in Theorem~1 of the main text and Theorem~\ref{theorem:overlaps_pool} above, it is sufficient to consider just the first $\tilde{L}$
	convolutional layers with unit strides, specifically $\tilde{L} = L$ for the case of a sequence of $K^d$ conv layers followed by global pooling
	(Fig.~\ref{fig:general_overlaps} with $P=1$), and $\tilde{L} = 1$ for the case of alternating $K^d$ conv and $2^d$ pooling layers
	(Fig.~\ref{fig:general_overlaps} with $P=2$). Under the above, the total receptive field of the $\tilde{L}$'th layer is simply
	$(K-1)\cdot\tilde{L} + 1$, which can be thought of a single large convolutional layer. Now, following the same proof sketch described above,
	we can use the combined convolutional layer to pair indices of $A$ and $B$ along the boundary between the two sets, where the size of the
	total receptive field determines the maximal number of pairs we can capture around each point on the boundary. In the special case of
	$K\tilde{L} > \alpha$, nearly any index of $A$ could be paired with a unique index from $B$. This results in a lower bound of
	$\min\{M,r_0,\ldots,r_{L-1}\}^{\Omega\left(\min\left\{\alpha^{d},K\tilde{L}\alpha^{d-1}\right\}\right)}$.
	
	\hfill $\square$ 
	
	\section{C. Entanglement Scaling in Deep-Recurrent Networks}
	
	In the following, we prove the result in Theorem~2 of the main text, regarding the entanglement scaling supported by deep RACs: 
	
	\emph{Proof} (of Theorem~2 of the main text).
	In~\cite{levine2017benefits}, a lower bound of $log\{\binom{\min\left\{R, M\right\}+\nicefrac{N}{2}-1}{\nicefrac{N}{2}}\}$ is shown for $A$ that is placed to the right of $B$ and  $\abs{A}=\abs{B}=\nicefrac{N}{2}$, for which the size of $\abs{A}$ is the largest possible under the conditions of Theorem~2 of the main text. 
	There, $R$ is the dimension of the RAC's hidden state.
	Essentially, the combinatorial dependence of the lower bound follows from the indistinguishability of duplicated indices.
	Given $\abs{A}\leq\abs{B}$, we designate the final $\abs{A}$ indices of the set $B$ to form a set $B^\ast$ which upholds by definition $\abs{B^\ast}=\abs{A}$. The lower bound in Theorem~2 of the main text is obtained by replacing $\nicefrac{N}{2}$ with $\abs{A}$ and continuing with the same exact proof procedure as in~\cite{levine2017benefits}, applied to $B^\ast$ and $A$, when all the residual initial $|B|-|A|$ indices, corresponding to the set $B\backslash B^\ast$, are kept fixed. Finally, for fixed $R,M$ the binomial term $\binom{\min\left\{R, M\right\}+\abs{A}-1}{\abs{A}}$ is polynomial in $\abs{A}$, therefore its logarithm obeys $\Omega\left(log\left\{\abs{A}\right\}\right)$.
	
	\hfill $\square$ 
	
	\bibliography{references_phys}
	
\end{document}